# Standoff Through-the-Wall Sensing at Ka-Band Microwave

S. Liao[1,2*], T. Elmer[1], S. Bakhtiari[1], N. Gopalsami[1], N. Cox[1], J. Wiencek[1], and A. C. Raptis[1]

[1]Nuclear Science and Engineering Division, Argonne National Laboratory, 9700 S. Cass Avenue, Argonne, IL, USA 60439

[2]Department of Electrical and Computer Engineering, Illinois Institute of Technology, 3300 S Federal St, Chicago, IL USA 60616

*Corresponding author, e-mail: sliao5@iit.edu

**ABSTRACT**

*Conventional microwave remote sensing/imaging of through-the-wall objects made of different materials is usually performed at frequencies below 3 GHz that provide relatively low spatial resolution. In this paper, we evaluate the ability and sensitivity of high-frequency microwave or millimeter wave standoff sensing of through-the-wall objects to achieve high spatial resolution. The target under study is a sandwich structure consisting of different object materials placed between two wall blocks. An Agilent® PNA-X series (model N5245A) vector network analyzer is used to sweep over the entire Ka-band (26.5 GHz to 40 GHz). The beam is then directed to a standard rectangular horn antenna and collimated by a 6-inch-diameter Gaussian lens towards the sandwich structure (wall block/object/wall block). The reflected electromagnetic wave is picked up by the same system as the complex S-parameter $S_{11}$. Both amplitude and phase of the reflected signal are used to recognize different materials sandwiched between the cement blocks. The experimental results are compared with the theoretical calculations, which show satisfactory agreement for the cases evaluated in this work.*

## INTRODUCTION

Compared to high-frequency electromagnetic waves such as millimeter wave (30-300 GHz [1-37]) and optics [38-48], the long-range and penetrative nature of microwave [49-61] through optically opaque dielectric materials makes it an ideal candidate for Through-the-Wall Imaging (TWI) of distant or inaccessible objects with the required spatial resolution. For such reason, recently, microwave standoff/remote TWI or sensing has attracted considerable amount of interest [62-69].

*Potential Applications*

The penetration property of the electromagnetic wave has found various industrial, civil and military applications [67, 70-75]. Various techniques at different frequency ranges have been extensively explored in the past decades. Most investigations were performed at the lower frequency range of microwaves, i.e., from DC to 3 GHz. This is because microwave attenuates less in this frequency range and in turn provides larger penetration depth [68]. What's more, there are commonly three operation modes: CW, FMCW and pulse. Table 1 provides some typical applications of TWI that are separated into 3 categories:

- Human inside Building

Through the wall searching and identification of human subjects is of great interest for many purposes, which include searching for 1) trapped human subjects after disasters such as fires, flooding, tornado etc.; 2) terrorists or hostages inside enclosed structures during counter-terrorism or rescue operations; and 3) soldiers in the battlefield. For example through-the-wall vital signs detection and monitoring using UWB (Ultra-Wide Band) radar signal has been investigated by [70] and X-band CW (Continuous Wave) radar has been studied in [67].

- Buried Humans

Searching for buried human is an important application of TWI. Typical examples are: 1) buried humans under rubbles after earthquakes; 2) buried humans under the snow after an avalanche; and 3) humans hidden or lost inside forest. For example, in [71], UHF-band and L-band CW radars are used to detect human subjects buried under simulated earthquake rubbles; also S-band radar is used to detect buried human subjects during snow avalanche in [72].

- Hidden Object Inspection

The penetration property of the electromagnetic wave can be used for noncontact inspection applications such as 1) hidden objects inside walls, 2) buried landmine, cables, and pipes etc., and 3) underground structures such as soil, rock stratigraphy, ice depth, and hidden tunnels for illegal immigration and transport of illicit materials. For example, [73] used an L-band FMCW (Frequency Modulated Continuous Wave) radar for detection of hidden objects inside walls, [74] investigated an S-band/C-band/Ku-band CW radar for detection of under-surface hidden objects, and [75] used a VHF airborne pulse radar for remote sensing of underground buried objects.

Table 1 Some Typical Applications of TWI

| Research Category | Year | Operation Modes | Frequency Band | Experiment or Theory |
|---|---|---|---|---|
| Humans inside Building | [70] | Pulse | X-band: 10 GHz | Experiment |
| | [67] | CW | | |
| Buried Humans | [71] | CW | UHF-band/L-band: 450 MHz/1150 MHz | Experiment |
| | [72] | | S-band: 2.42 GHz | |
| Hidden Object Inspection | [73] | FMCW | L-band: 1-2 GHz | Experiment |
| | [74] | CW | S-band/C-band/Ku-band: 3.6-4.0 GHz/5.8-6.8 GHz/14-15 GHz | Experiment |
| | [75] | Pulse | VHF: 50 MHz/100 MHz | Simulation & Experiment |

***Contribution of This Paper***

One disadvantage of low frequency microwaves is its relatively low spatial resolution limited by the diffraction effect of electromagnetic waves. Higher frequency microwave techniques have also been investigated by other researchers primarily for nondestructive evaluation (NDE) of layered materials ([76-80]). Those studies on the other hand have been limited mostly to measurements over a relatively narrow band. In this paper, we investigate TWI at the upper end of microwave and the lower end of millimeter wave (mmW) spectrum covered by the Ka-band (26.5 GHz to 40 GHz), aimed at achieving higher spatial resolution. Experimental sensing of various materials sandwiched between two cement blocks in the Ka band is compared with the theoretical calculations to validate the feasibility of standoff through-the-wall sensing/imaging ability at such a high frequency range.

**EXPERIMENTAL SETUP**

The experimental setup used to investigate the concept of mmW TWI is shown in Fig. 1. The microwave apparatus consists of a 4-port Agilent® PNA-X series (model N5245A) vector network analyzer (VNA) for stepping through frequencies from 26.5 GHz to 40 GHz, a Ka-band standard rectangular horn antenna, and a 6-inch-diameter Gaussian horn antenna for collimating the beam. The plane wave propagates towards a sandwiched structure which is perpendicular to the beam path and is located ~30 inches away from the lens. The reflected wave is picked up by the same Ka-band horn antenna and is processed by the VNA to determine the complex $S_{11}$ parameter (reflection coefficient) The sandwich structure is composed of two wall blocks (1.4-inch thick cement) with a 7-inch spacing where different materials can be placed. A LabVIEW® program running on a computer was developed for recording of the $S_{11}$ parameter and for further processing.

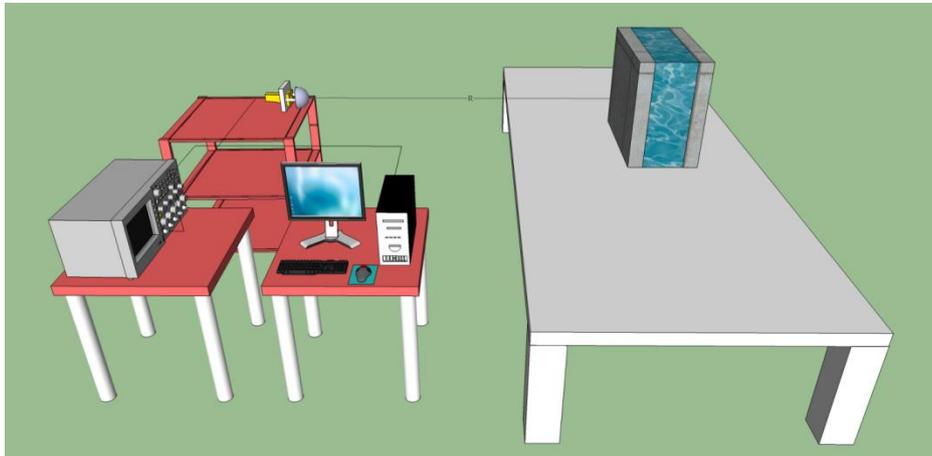

*Fig. 1 Experimental setup of the standoff through-the-wall sensing of different objects: water, soil, rocks, and metal plate etc. The object is sandwiched between two wall blocks to form the sandwich structure. R is the range of the measurement.*

**THEORETICAL SIMULATION**

Figure 2 depicts the simplified numerical model of the experimental setup shown in Fig. 1. Without loss of generality, in the far field region the collimated beam can be approximated by an incident plane wave which reduces the model to plane wave incidence upon a layered dielectric structure. The problem may then be readily solved through the boundary matching method on all four boundaries separated by different materials. In reference to Fig. 2, from left to right, boundary #1 is air/wall-block boundary, boundary #2 is wall-block/object boundary, boundary #3 is object/wall-block boundary, and boundary #4 is wall-block/air boundary. In reference to Table 1, the electric field (E-field) and magnetic field (H-field) inside each of the three layers of the target structure can be expressed as forward and backward propagating waves [81], where $\gamma_w$ and $\gamma_o$ are wave propagation constants that include the attenuation effect inside the wall block and inside the object, respectively. As shown in Table 2, the boundary conditions at each interface may then be obtained by enforcing the continuity of both E-field and H-field obtained, where $k$ is the wave vector in the air, $t_w$ and $t_0$ are the thickness of the wall block and the object, respectively; $\gamma_w$ and $\gamma_0$ are the propagation constant of the wall block and the object, respectively.

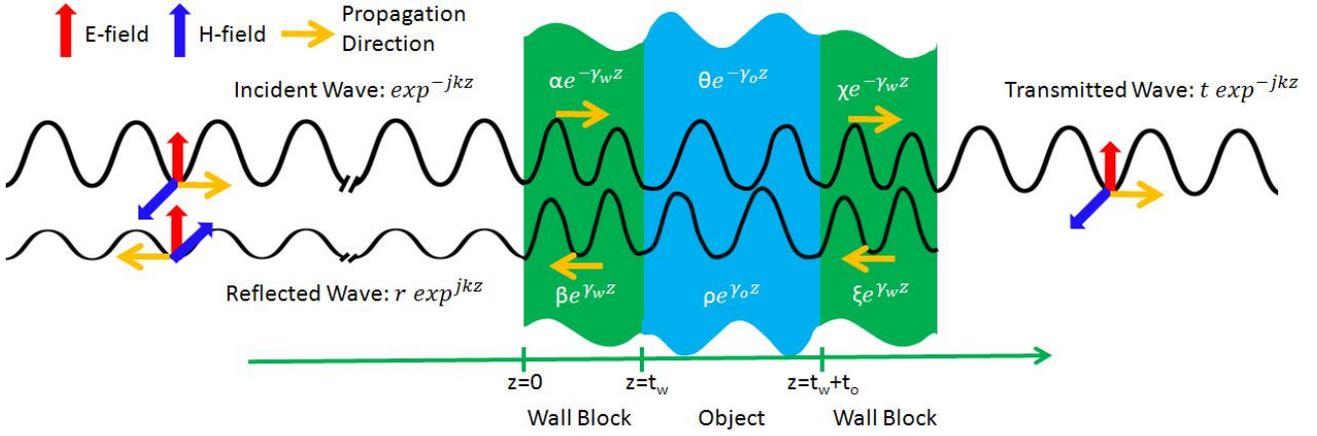

Fig. 2 Simplified model of plane wave propagation through the wall-block/object/wall-block sandwich structure.

**Table 1. Total E/H-field Inside Each Region**

| Regions | #1: Air | #2: Wall Block | #3: Object | #4: Wall Block | #5: Air |
|---|---|---|---|---|---|
| E-field | $e^{-jkz} + re^{jkz}$ | $\alpha e^{-\gamma_w z} + \beta e^{\gamma_w z}$ | $\theta e^{-\gamma_o z} + \rho e^{\gamma_o z}$ | $\chi e^{-\gamma_w z} + \xi e^{\gamma_w z}$ | $te^{-jkz}$ |
| H-field × jωμ | $jk(e^{-jkz} - re^{jkz})$ | $\gamma_w(\alpha e^{-\gamma_w z} - \beta e^{\gamma_w z})$ | $\gamma_o(\theta e^{-\gamma_o z} - \rho e^{\gamma_o z})$ | $\gamma_w(\chi e^{-\gamma_w z} - \xi e^{\gamma_w z})$ | $jkte^{-jkz}$ |

**Table 2. E/H-field Boundary Conditions on Each Boundary**

| Boundaries | #1: Air/Wall Block | #2: Wall Block/Object | #3: Object/Wall Block | #4: Wall Block/Air |
|---|---|---|---|---|
| E-field Boundary Condition | $1 + r = \alpha + \beta$ | $\alpha e^{-\gamma_w t_w} + \beta e^{\gamma_w t_w} = \theta e^{-\gamma_o t_w} + \rho e^{\gamma_o t_w}$ | $\theta e^{-\gamma_o [t_w+t_o]} + \rho e^{\gamma_o [t_w+t_o]} = \chi e^{-\gamma_w [t_w+t_o]} + \xi e^{\gamma_w [t_w+t_o]}$ | $\chi e^{-\gamma_w [2t_w+t_o]} + \xi e^{\gamma_w [2t_w+t_o]} = te^{-jk[2t_w+t_o]}$ |
| H-field Boundary Condition | $jk(1-r) = \gamma_w(\alpha - \beta)$ | $\gamma_w(\alpha e^{-\gamma_w t_w} - \beta e^{\gamma_w t_w}) = \gamma_o(\theta e^{-\gamma_o t_w} - \rho e^{\gamma_o t_w})$ | $\gamma_o(\theta e^{-\gamma_o [t_w+t_o]} - \rho e^{\gamma_o [t_w+t_o]}) = \gamma_w(\chi e^{-\gamma_w [t_w+t_o]} - \xi e^{\gamma_w [t_w+t_o]})$ | $\gamma_w(\chi e^{-\gamma_w [2t_w+t_o]} - \xi e^{\gamma_w [2t_w+t_o]}) = jkte^{-jk[2t_w+t_o]}$ |

The boundary conditions in Table 2 can be readily solved in MATLAB® for all eight unknowns (i.e., $r, \alpha, \beta, \theta, \rho, \chi, \xi,$ and $t$) since we have eight boundary conditions. In matrix form, Table 2 reduces to,

$$\overline{\overline{M}}(f, \varepsilon_w, \varepsilon_o, t_w, t_o) \times \overline{u}(\alpha, \beta, \theta, \rho, \chi, \xi) = [1; jk; 0; 0; 0; 0; 0; 0]^T, \tag{1}$$

$$\overline{\overline{M}}(f, \varepsilon_w, \varepsilon_o, t_w, t_o) = \begin{bmatrix} -1 & 1 & 1 & 0 & 0 & 0 & 0 & 0 \\ jk & \gamma_w & -\gamma_w & 0 & 0 & 0 & 0 & 0 \\ 0 & e^{-\gamma_w t_w} & e^{\gamma_w t_w} & -e^{-\gamma_o t_w} & -e^{\gamma_o t_w} & 0 & 0 & 0 \\ 0 & \gamma_w e^{-\gamma_w t_w} & -\gamma_w e^{-\gamma_w t_w} & -\gamma_o e^{-\gamma_o t_w} & \gamma_o e^{-\gamma_o t_w} & 0 & 0 & 0 \\ 0 & 0 & 0 & e^{-\gamma_o [t_w+t_o]} & e^{\gamma_o [t_w+t_o]} & -e^{-\gamma_w [t_w+t_o]} & -e^{-\gamma_w [t_w+t_o]} & 0 \\ 0 & 0 & 0 & \gamma_o e^{-\gamma_o [t_w+t_o]} & -\gamma_o e^{\gamma_o [t_w+t_o]} & -\gamma_w e^{-\gamma_w [t_w+t_o]} & \gamma_w e^{\gamma_w [t_w+t_o]} & 0 \\ 0 & 0 & 0 & 0 & 0 & e^{-\gamma_w [2t_w+t_o]} & e^{\gamma_w [2t_w+t_o]} & e^{-jk[2t_w+t_o]} \\ 0 & 0 & 0 & 0 & 0 & \gamma_w e^{-\gamma_w [2t_w+t_o]} & -\gamma_w e^{\gamma_w [2t_w+t_o]} & -jke^{-jk[2t_w+t_o]} \end{bmatrix},$$

where $f$ is the frequency, and $\varepsilon_w, \varepsilon_o$ are the dielectric constant of the wall block and the object, respectively. The calculated results and their comparison with the experimental results will be discussed in the following sections.

**EXPERIMENTAL RESULT**

In our experiment, 1601 frequencies evenly distributed between 26.5 GHz and 40 GHz were recorded for all the measurement scenarios. The IF-bandwidth of the VNA was set at 10 Hz to increase the Signal to Noise Ratio (SNR), which is inversely proportional to the IF-bandwidth.

*System calibration*

Due to the existence of background stray field $E_b$, we needed to calibrate our measurement system. The measured S-parameter $S_{11}$ can be expressed as follows [57],

$$S_{11} = (E_b + E_r)T(f) = E_i(r_b + r)T(f), \tag{2}$$

where $E_b$, $E_r$, and $E_i$ are the background stray field, the reflected field due to the structure and the incident field, respectively, and $T(f)$ is the transfer function of the system. The goal is to ultimately extract the true reflection signal $r$ from the measurements. The S-parameter for the background $S_{11}^b$ and the S-parameter for the reflection from a metal surface $S_{11}^m$, which are measurable quantities that can be expressed as follows,

$$S_{11}^b = E_i r_b T(f), \tag{3}$$
$$S_{11}^m = E_i(r_b - 1)T(f), \tag{4}$$

from which one gets,

$$E_i T(f) = S_{11}^b - S_{11}^m. \qquad (5)$$

By combining Eq. (2) and Eq. (5), the true reflection coefficient from the structure is then given by,

$$r = \frac{S_{11} - S_{11}^b}{S_{11}^b - S_{11}^m}. \qquad (6)$$

### *Dielectric constant of wall block*
In order to solve Eq. (1), the dielectric constant of the wall block has to be first determined. This was obtained through direct measurement of the S-parameter $S_{11}$ at 101 frequency points within the Ka band for an individual wall block. The measurement data may then be fitted by using the following theoretical formula [57],

$$r_w = \frac{\left[\frac{1}{\varepsilon_w} - 1\right]\left[1 - \exp(-j2k_w t_w)\right]}{\left[\frac{1}{\sqrt{\varepsilon_w}} + 1\right]^2 - \left[\frac{1}{\sqrt{\varepsilon_w}} - 1\right]^2 \exp(-j2k_w t_w)}. \qquad (7)$$

By using this approach, we obtained an average value of $\varepsilon_w = 12.4(1 - j0.003)$ for the dielectric constant of the wall block. Comparison of the experimental result with that calculated using Eq. (7) is shown in Fig. 3.

### *$S_{11}$ of the sandwich structure*
The S-parameter $S_{11}$ for sandwiched water object is shown in Fig. 4, together with the theoretical results calculated according to Eq. (1). A snapshot of data in Fig. 3 over a narrow frequency range is shown in Fig. 5. The parameters used for fitting are $\varepsilon_w = 12.4(1 - j0.003)$, obtained from Fig. 3, and water, as the sandwiched object, with a dielectric constant $\varepsilon_o = 77$. The sandwich structure was approximately 30 inches ($R \sim 30''$) away from the antenna.

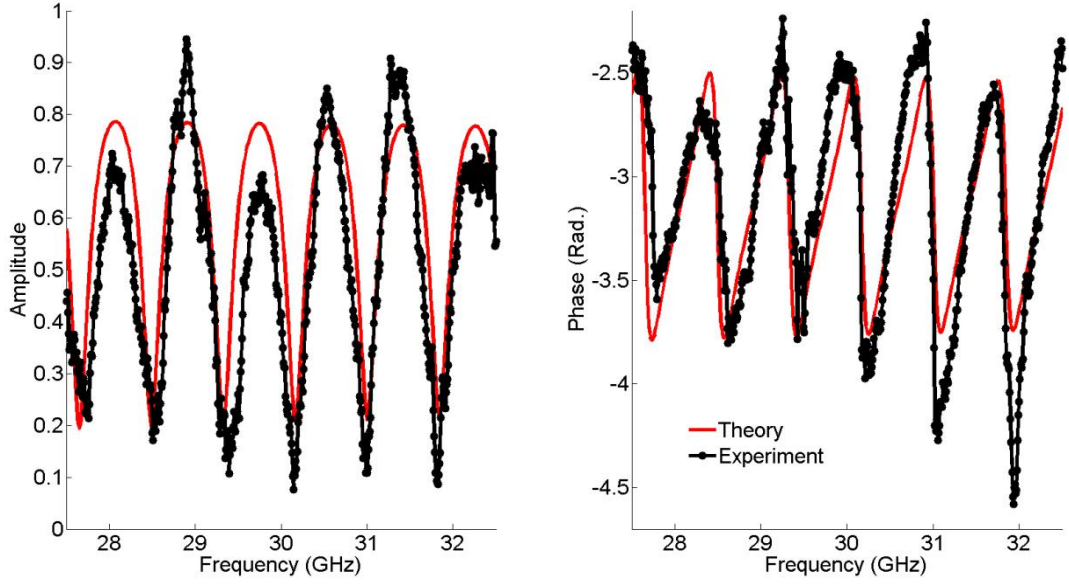

*Fig. 3 Experimental reflection is fitted to theoretical calculation, i.e., Eq. (7), to obtain the dielectric constant of the wall block used in the experiment, which was determined to be $\varepsilon_w = 12.4(1 - j0.003)$.*

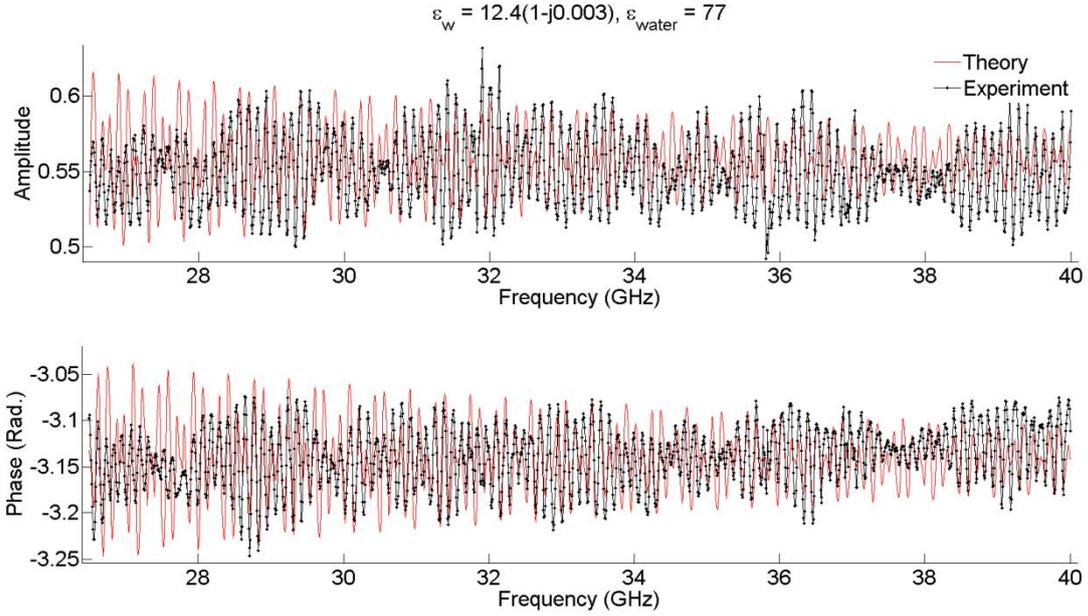

*Fig. 4 Experimental and calculated S-parameter S11 for wall block/water/wall block sandwich structure where $\varepsilon_w = 12.4(1 - j0.003)$ and $\varepsilon_o = 77$ for water.*

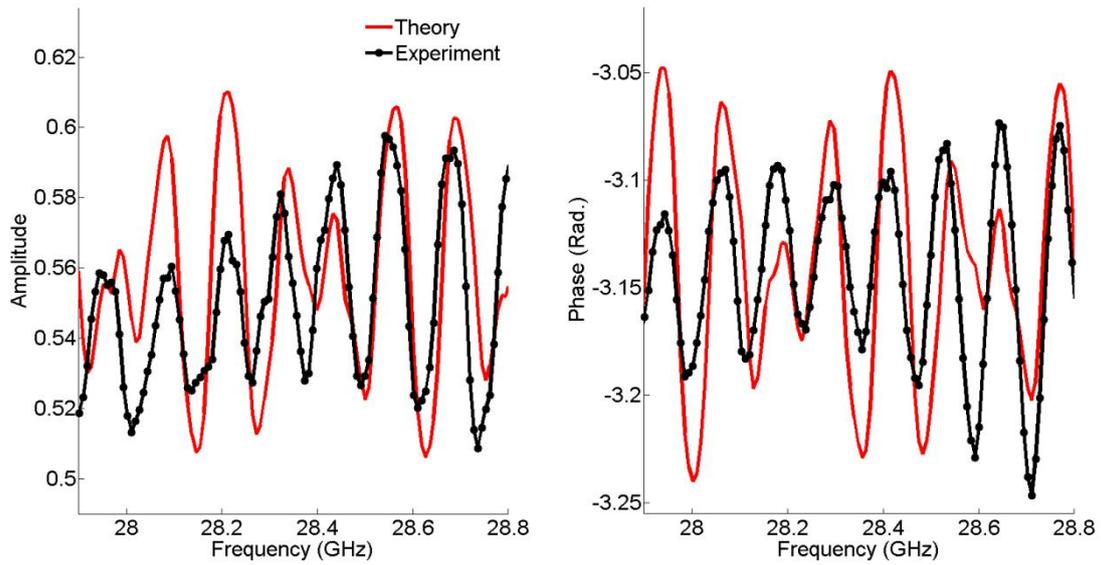

*Fig. 5 A snapshot of Fig. 3 to show the details.*

Following the validation of the procedure for extraction of dielectric properties of test object, several experiments were carried out by using different materials sandwiched between the two wall blocks, i.e., soil, rocks and metal. To demonstrate the ability to see the contrast between different objects behind the wall, the difference (with that of water as the object) of the S-parameter, $\Delta S_{11} \equiv S_{11} - S_{11}(water)$, for three different materials is shown in Fig. 6. Both the amplitude and the phase spectrum are displayed. Once again, Fig. 7 provides a snapshot of Fig. 6, allowing for more detailed observation of the data over a narrow bandwidth. From the S-parameter contrast, one can clearly distinguish different materials by combining both the amplitude and the phase information.

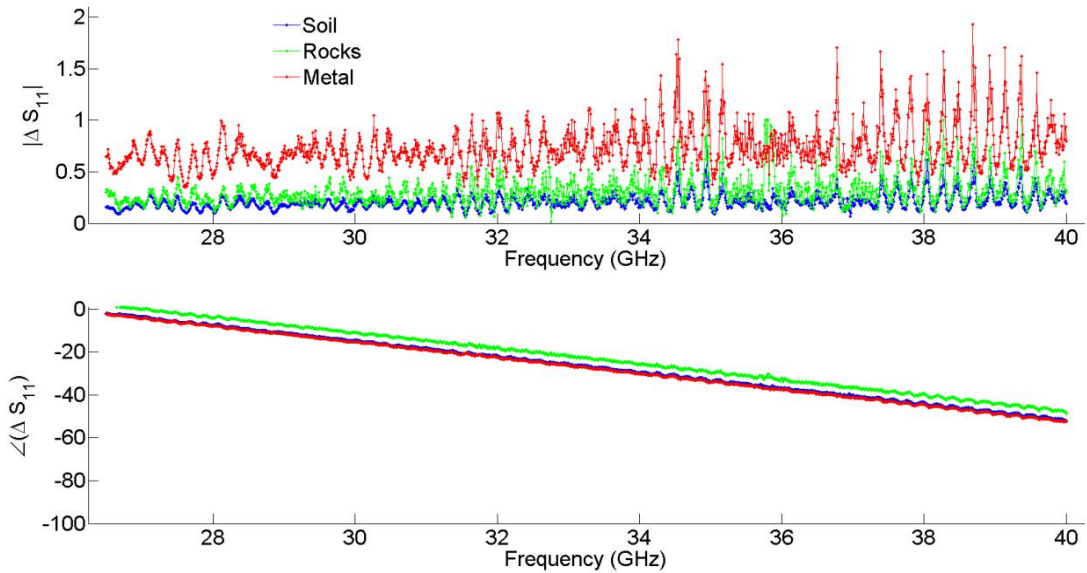

*Fig. 6 Experimental S-parameter contrast $\Delta S_{11} \equiv S_{11} - S_{11}(water)$ for different materials in the middle layer of the sandwich structure, i.e., soil, rocks and metal.*

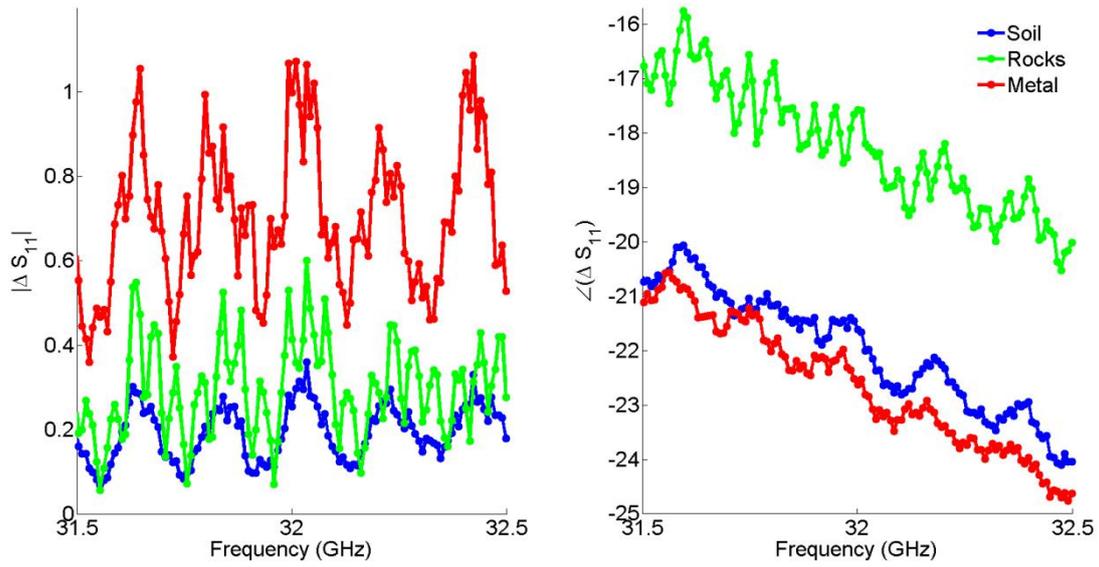

*Fig. 7 A snapshot of Fig. 6 showing the details over a narrow band.*

**DISCUSSION**

We have demonstrated the feasibility of standoff through-the-wall sensing of different materials at Ka band using a wide-band stepped-frequency measurement approach. Both theoretical and experimental results show promise for application of this technique to TWI at these relatively high microwave frequencies for high resolution imaging of hidden objects (e.g., covered by wall, soil, snow, etc.). The detection range can be estimated by Friis transmission equation,

$$P_d(\min.) = P_t G^2 \left(\frac{\lambda}{4\pi R}\right)^2 \Rightarrow R = \lambda \frac{G}{4\pi} \sqrt{\frac{P_t}{P_d(\min.)}} \; , \qquad (8)$$

where $R$ is the range; $P_t$ and $P_d(\min.)$ are the transmitted and the minimum detectable power by the system, respectively, $\lambda$ is the operating wavelength and $G$ is the gain of the system.

Future extension of this monostatic setup to mobile systems with multiple antennas (array configuration) is expected to provide even higher resolution for 2D TWI applications. If combined with a frequency sweeping function, 3D TWI is also possible, allowing identification of objects behind a wall with known dielectric constant and thickness [Ahmad, 2008].

**CONCLUSION**

Through-the-wall sensing of various materials at Ka band have been investigated using a monostatic stepped-frequency experimental setup. A network analyzer was used to measure S-parameter $S_{11}$ of an object within a layered structure over the frequency range of 26.5 GHz to 40 GHz. The target in this work was a wall-block/object/wall-block sandwich structure. The hidden objects in the middle layer included water, soil, rocks, and metal. Combination of the amplitude and the phase information of the reflected signal shows clear contrast among various objects sandwiched between two wall blocks. Theoretical calculations based on a plane wave solution obtained using a boundary matching technique were also performed. Numerical results showed good agreement with that of the experimental data. The results of this investigation suggest that high-resolution mmW through-the-wall sensing and imaging techniques could provide a viable alternative to low-frequency microwave TWI.